\documentclass[10pt,twocolumn]{article}
\usepackage[margin=0.75in]{geometry}
\usepackage{amsmath, amssymb, mathtools}
\usepackage{graphicx}
\usepackage{booktabs}
\usepackage{multirow}
\usepackage[colorlinks=true,allcolors=blue]{hyperref}
\usepackage{cleveref}
\usepackage[table]{xcolor}
\usepackage[T1]{fontenc}
\usepackage[utf8]{inputenc}
\usepackage{microtype}
\usepackage{cite}
\usepackage{authblk}
\usepackage{mathptmx} 

\let\origthebibliography\thebibliography
\renewcommand\thebibliography[1]{%
  \origthebibliography{#1}%
  \setlength{\itemsep}{0pt}%
  \setlength{\parsep}{0pt}%
  \setlength{\topsep}{2pt}}

\title{\textbf{MedSAM2-Anatomy: Training-Free Inference-Time Optimization for Musculoskeletal Segmentation}}
\author[1]{John~Garcia-Henao\thanks{This work was supported in part by the Digitalization Initiative of the Zurich Higher Education Institutions (DIZH) and Balgrist University Hospital through the MedTwins Agil.IT and MIRO Suite projects. Corresponding author: \texttt{john.garciahenao@balgrist.ch}}}
\author[1]{Nicholas~B\"unger}
\author[1]{Benedikt~Herzog}
\author[2]{Cindy~Guerrero~Toro}
\author[3]{Benjamin~Vella}
\author[4]{Matthias~Biner}
\author[4]{Rico~Br\"utsch}
\author[4]{Carmen~Castroviejo~Fernandez}
\author[4]{Felix~\"Ottl}
\author[2]{Norman~Juchler}
\author[3]{Armando~Hoch}
\author[4]{Bettina~Hochreiter}
\author[2]{Sven~Hirsch}
\author[1]{Sebastiano~Caprara}

\affil[1]{Digital Medicine Unit Team, Balgrist University Hospital, 8008 Zurich, Switzerland}
\affil[2]{Research Centre for Computational Health, Zurich University of Applied Sciences (ZHAW), 8820 W\"adenswil, Switzerland}
\affil[3]{Hip and Pelvis Surgery Team, Balgrist University Hospital, 8008 Zurich, Switzerland}
\affil[4]{Shoulder and Elbow Surgery Team, Balgrist University Hospital, 8008 Zurich, Switzerland}

\date{}

\begin{document}
\maketitle
\begin{abstract}
High-resolution 3D segmentation of hip and shoulder anatomy from CT and MRI is essential for surgical planning, yet frozen segmentation models often fail under domain shift. CNN-based expert models are fully automatic but lack adaptability, whereas promptable foundation models generalize better but require manual prompting. We present MedSAM2-Anatomy, a training-free inference-time optimization framework that improves frozen segmentation models without retraining or human interaction. A frozen expert model generates anatomical priors that are automatically converted into multiple prompt hypotheses for a frozen 3D foundation model. Candidate masks are fused while anatomically implausible priors are rejected. No model weights are updated and no manual prompts are required. TotalSegmentator and MedSAM2 are used as representative expert and foundation models, allowing the contribution of the inference policy to be isolated. Evaluation on the independent Balgrist-V0 CT and MRI cohorts shows that inference-time optimization increases median Dice from $0.71$ to $0.92$ on hip MRI and from $0.89$ to $0.92$ on shoulder CT, while reducing median HD95 on hip MRI from $22.0$\,mm to $5.0$\,mm. On public TotalSegmentator benchmarks, the expert model remains strongest, indicating that the optimal fusion strategy depends on the reliability of the expert prior. These results demonstrate that training-free inference-time optimization provides a practical strategy for improving frozen segmentation models without manual prompting.
\end{abstract}

\section{Introduction}
\label{sec:introduction}
Surgical planning and morphological analysis of the hip and shoulder rely on three-dimensional models of bone anatomy reconstructed from CT and MRI. Femoroacetabular impingement (FAI) assessment requires the acetabulum and the proximal femur; posterior shoulder instability (PSI) planning requires the scapula and the humerus. These models are currently produced by manual or semi-automatic contouring of high-resolution volumes, which is slow, operator-dependent, and the main bottleneck between image acquisition and a usable plan. Automatic multi-label segmentation removes that bottleneck only if it is accurate enough on the institutional imaging protocol actually used in theatre.

Anatomy-specific convolutional neural networks are the established solution. U-Net \cite{unet}, nnU-Net \cite{nnunet}, and TotalSegmentator \cite{totalsegmentator,totalsegmentator_mri} are fully automatic, require no user input, and are accurate within their training distribution. Their weakness is that their parameters are fixed once training ends. On acquisition protocols, resolutions, or pathologies that differ from the training data, delineation degrades, and the only established remedy is to retrain or fine-tune on annotated target-domain data, which is precisely the annotation cost the model was meant to avoid.

Promptable foundation models behave differently. SAM \cite{sam}, SAM2 \cite{sam2}, and their medical adaptations MedSAM \cite{medsam}, MedSAM2 \cite{medsam2}, SIT-SAM \cite{sitsam}, and SegmentAnyBone \cite{segmentanybone} transfer across modalities and anatomies without task-specific training \cite{generalist}. They shift the difficulty rather than remove it: for identical weights and an identical input volume, moving a prompt by a few slices can change the segmentation substantially, and in 3D the choice of which slices to seed is itself a design decision \cite{prompt_sensitivity}.

In practice this decision is delegated to a user. Interactive systems such as nnInteractive \cite{nninteractive} are explicitly built around that assumption and are effective, but they make segmentation quality a function of human effort. For batch processing of an institutional cohort, or for an automated planning pipeline, a human in the loop is not a solution but the constraint to be removed.

The two families therefore fail in complementary ways: convolutional experts are automatic but brittle under domain shift, whereas foundation models are adaptable but require a human to place the prompt. Both leave the same resource unused, namely the freedom that remains in how a volume is routed through an already-trained model. This is the resource the present work exploits, and it frames our central question: \emph{how can we systematically optimise frozen state-of-the-art segmentation models at inference time, without retraining or requiring human interaction?}

Throughout this paper, \emph{inference-time optimisation} denotes the optimisation of the \emph{execution policy} of frozen models rather than of their parameters. The policy comprises four components: automatic prompt generation from an anatomical prior, prompt sampling, that is how many hypotheses are explored and where, the fusion strategy that combines the resulting candidate masks, and anatomical plausibility filtering, which decides whether a prior is admitted to the fusion at all. These choices are selected once on a target-domain cohort and then applied unchanged to every case; no gradient is computed and no weight is modified at any stage.

MedSAM2-Anatomy is the concrete instance we study. It treats a frozen expert CNN not as a predictor but as a prompt generator, samples several prompt hypotheses from its output, propagates each through a frozen 3D foundation model, and aggregates the resulting candidates into a single mask. A plausibility test on the expert prior discards priors that cannot correspond to real anatomy before they influence the result. It is not a new segmentation architecture. Its purpose is to make the composition of frozen models a tunable object and to determine which parts of that composition matter.

Two consequences of this framing govern how the study is designed. First, the objective is not to demonstrate superiority over all available segmentation methods, but to test whether inference-time optimisation systematically improves the frozen models it is given. TotalSegmentator and MedSAM2 instantiate the expert and foundation roles because they are representative, widely used, and publicly released; holding both fixed is what allows the measured differences to be attributed to the inference policy rather than to a stronger network. Second, because the policy is selected on a target domain, that domain carries the evidential weight: primary evaluation is performed on the institutional Balgrist-V0 CT and MRI cohorts with independent expert manual contours, while the public TotalSegmentator CT and MRI test sets provide secondary benchmark evaluation.

Our contributions are:
\begin{enumerate}
    \item An inference-time optimisation framework that couples a frozen expert CNN with a frozen promptable 3D foundation model through automatic prompt generation, multi-hypothesis propagation, mask fusion, and anatomical plausibility filtering, with no training, fine-tuning, or manual prompting.
    \item An ablation on the institutional Balgrist-V0 cohorts that identifies the fusion layer as the inference-time component governing final segmentation quality, and selects union fusion with rejection of implausible priors over softer aggregation rules, additional prompts, and post-hoc mask repair.
    \item A four-cohort evaluation for hip and shoulder anatomy on CT and MRI, with the target domain as the primary evaluation and two public test sets as a secondary benchmark check, which quantifies when inference-time optimisation helps and when it does not, and shows that the best fusion strategy depends on the reliability of the expert prior.
\end{enumerate}

\section{Inference-Time Optimization Framework}
\label{sec:method}

\subsection{Problem Setting and Constraints}
Let $X \in \mathbb{R}^{D \times H \times W}$ be a 3D volume acquired with modality $m \in \{\text{CT}, \text{MRI}\}$ and let $A = \{a_1, \dots, a_K\}$ be the target anatomies. The objective is a multi-label mask $\hat{Y} \in \{0, \dots, K\}^{D \times H \times W}$.

The framework is defined by what it is not allowed to do. No parameter of any model is updated, no gradient is computed, and no prompt is supplied by a user. Both constituent models are used exactly as released. All degrees of freedom therefore live outside the networks, in the execution policy that routes the volume through them. That policy has four tunable components: (i) automatic prompt generation from the anatomical prior, (ii) prompt sampling, that is the number and placement of prompt hypotheses, (iii) the fusion strategy that combines the resulting candidates, and (iv) anatomical plausibility filtering, which decides whether a prior is admitted at all. Inference-time optimisation means selecting among these components, and the selection is made once on a target-domain cohort rather than per case.

The expert and foundation models are held fixed and instantiated with representative public checkpoints. This is deliberate: any measured difference between the composition and its individual parts is then attributable to the execution policy and not to a change of network.

\begin{figure*}[t]
    \centering
    \includegraphics[width=\textwidth]{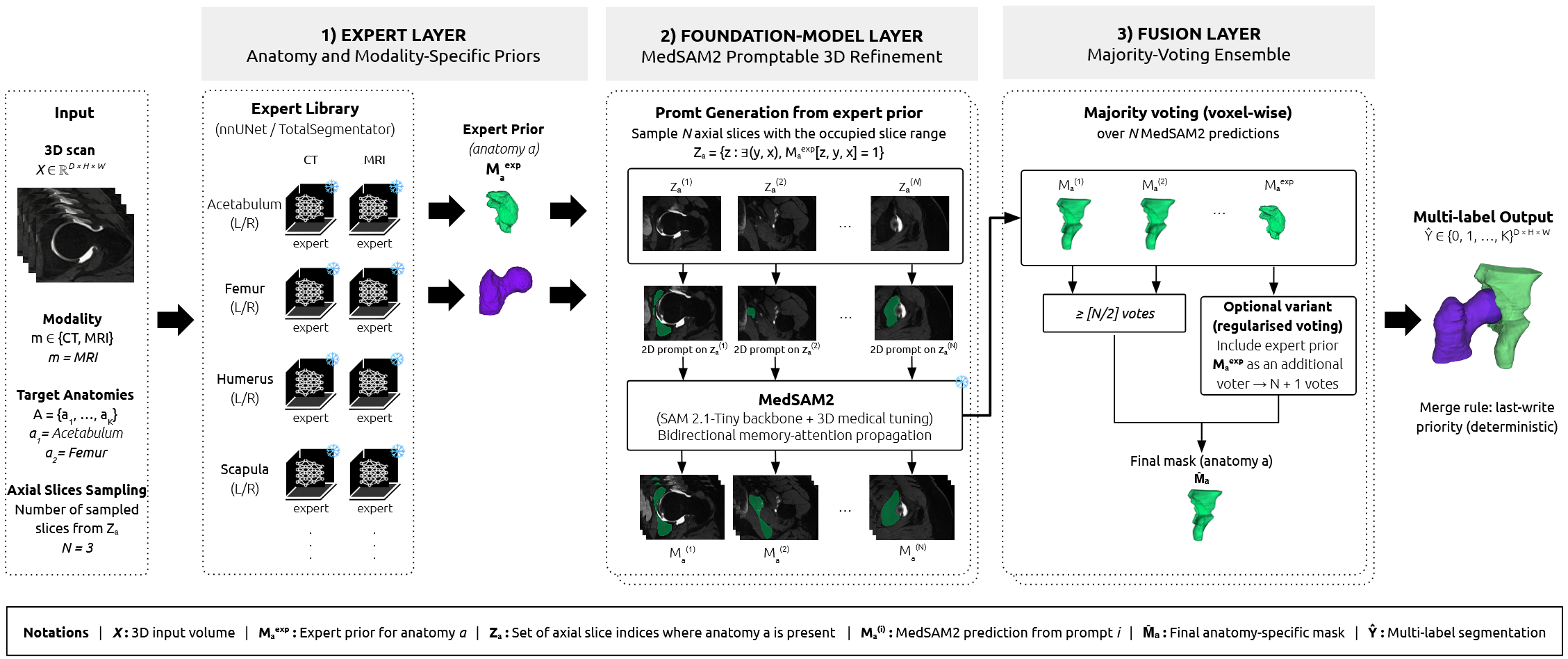}
    \caption{MedSAM2-Anatomy as an inference-time optimisation pipeline. A frozen expert network produces the prior $M^{exp}_a$, automatic prompt sampling turns that prior into $N$ prompt hypotheses, a frozen foundation model propagates each hypothesis into a candidate mask $\{M^{(i)}_a\}$, and the fusion layer aggregates the candidates into $\hat{M}_a$ before multi-label merging into $\hat{Y}$. All weights remain frozen; the tunable choices are the prompt budget, the fusion strategy, and the plausibility filter.}
    \label{fig:arch}
\end{figure*}

\subsection{Pipeline Overview}
The pipeline has five stages, all executed at inference time. (i) A frozen expert network segments the volume and provides a coarse anatomical prior. (ii) The prior is converted automatically into $N$ prompt hypotheses distributed along the structure. (iii) A frozen promptable foundation model propagates each hypothesis through the volume, producing $N$ candidate masks. (iv) The candidates are fused into one binary mask per anatomy. (v) Priors that fail an anatomical plausibility test are rejected so that they never reach the fusion. The stages are shown in \cref{fig:arch}.

The two models play complementary roles. The expert is used only for localisation, which is the property that survives domain shift; the foundation model is used only for delineation, which is where the expert degrades. Neither is asked to do the part it is bad at.

\subsection{Stage 1: Expert Prior Generation}
A deterministic router maps the volume to a binary prior for each anatomy $a \in A$,
\begin{equation}
    M^{exp}_a = f_{\text{exp}}(X, a),
    \label{eq:expert}
\end{equation}
where $f_{\text{exp}}$ is instantiated with the frozen TotalSegmentator \cite{totalsegmentator} network,
\begin{equation}
    f_{\text{exp}}(X, a) \equiv \text{TotalSegmentator}_a(X).
    \label{eq:prior}
\end{equation}
The prior is a spatial conditioner, not a candidate output. Its boundaries are never used directly; only its position and extent enter the pipeline. The expert runs once per scan regardless of $K$.

\subsection{Stage 2: Automatic Multi-Prompt Sampling}
Prompts are derived from the prior rather than from a user. We first take the extent of the prior along the longitudinal axis,
\begin{equation}
    [z_{\min}, z_{\max}] = \text{BoundingBox}_z(M^{exp}_a),
    \label{eq:zrange}
\end{equation}
and place $N$ evenly spaced seed slices inside it,
\begin{equation}
    S = \{z_i\}_{i=1}^{N}, \qquad
    z_i = \left\lfloor z_{\min} + \frac{i\,(z_{\max}-z_{\min})}{N+1} \right\rfloor .
    \label{eq:sample}
\end{equation}
The seeds spread over the structure while avoiding its two end slices, where the cross-section is smallest and the prior least reliable. Sampling several seeds rather than one is what turns prompt selection from a fixed guess into a set of hypotheses to be explored: each $z_i$ is an independent bet on where the structure is best characterised, and the fusion layer decides afterwards which parts of those bets to keep. The placement is deterministic, so a configuration can be re-run bit-for-bit.

\subsection{Stage 3: Frozen Foundation-Model Propagation}
For each seed slice $z_i$ we extract 2D mask and bounding-box prompts $\mathcal{P}(M^{exp}_a, z_i)$ from the prior and pass them to the frozen MedSAM2 \cite{medsam2} model, which propagates the segmentation through the volume with bidirectional memory attention:
\begin{equation}
    M^{(i)}_a = \text{MedSAM2}(X, \mathcal{P}(M^{exp}_a, z_i)).
    \label{eq:candidate}
\end{equation}
This produces $N$ candidate masks per anatomy, each representing a complete 3D segmentation generated from a different seed slice. Their variability provides a per-case measure of sensitivity to prompt-slice selection. An optional largest connected-component filter may be applied to each candidate. Computational cost scales linearly with $N$ for each anatomy.

\subsection{Stage 4: Fusion}
\label{sec:fusion}
The fusion layer converts the $N$ candidates into one mask and is the inference-time component that determines final segmentation quality; \cref{sec:ablation} quantifies this. Candidates are aggregated voxel-wise by thresholding the number of propagations that claim a voxel:
\begin{equation}
    \hat{M}_a = \mathbb{I}\left(\sum_{i=1}^N M^{(i)}_a \geq \tau\right),
    \label{eq:fuse}
\end{equation}
where $\mathbb{I}(\cdot)$ is the indicator function and $\tau \in \{1,\dots,N\}$ sets the required consensus. We use the union rule $\tau = 1$. The majority rule $\tau = \lceil N/2 \rceil$ is the natural default, but requiring a voxel to be claimed by half the propagations before it survives erodes boundaries and systematically under-segments thin structures; \cref{sec:ablation} quantifies the cost. Fixing $\tau$ rather than tying it to $N$ also keeps the consensus comparable when $N$ is varied, since $\lceil N/2 \rceil / N$ is not monotonic in $N$.

\Cref{eq:fuse} binarises each pass before counting. The alternative is to fuse the raw logit volumes $L^{(i)}_a$ and threshold once,
\begin{equation}
    \hat{M}^{\mathrm{soft}}_a = \mathbb{I}\!\left(
        \frac{\sum_i w_i(z)\, L^{(i)}_a}{\sum_i w_i(z)} > 0 \right),
    \quad
    w_i(z) = e^{-\left(\frac{z - z_i}{\sigma}\right)^2},
    \label{eq:soft}
\end{equation}
which preserves the distinction between a pass that is confidently empty and one that is merely uncertain because the slice lies far from its seed $z_i$. Uniform weights recover the plain logit mean, while a finite $\sigma$ lets the nearest-seeded pass dominate each slice. Two further operations act after fusion rather than during it: the largest-connected-component filter may be applied to the fused mask instead of to each pass, and enclosed cavities may be filled and single-slice dropouts bridged by 3D morphological closing. \Cref{sec:ablation} evaluates all of these against \cref{eq:fuse}; none is retained.

\subsection{Stage 5: Anatomical Plausibility Filtering}
\label{sec:guard}
Union fusion retains every voxel any pass observed, which recovers the volume that voting removes but exposes a failure mode of the expert layer. When the field of view truncates one side of a bilateral structure, the prior for the unscanned side is returned as a small sliver rather than as nothing. A track seeded inside that sliver has no true structure to lock onto and drifts onto the contralateral anatomy, and the union then keeps the drifted voxels.

We therefore gate each bilateral prior on a laterality plausibility test. Writing $\bar{a}$ for the contralateral partner of $a$, the prior is used only if
\begin{equation}
    \big| M^{exp}_a \big| \;\geq\; \rho \, \big| M^{exp}_{\bar{a}} \big|,
    \qquad \rho = 0.15,
    \label{eq:guard}
\end{equation}
and the anatomy is skipped otherwise. The test is evaluated before any prompting, so a rejected prior generates no hypothesis and contributes nothing to \cref{eq:fuse}. The threshold is not sensitive: observed slivers occupy $0.0003$--$0.047$ of their partner while genuine pairs on whole-body scans sit near unity, so $\rho$ falls in an empty region of the distribution. Being a relative test, it is vacuous when only one side of a pair is requested or when the partner prior is empty; it targets specifically the sliver-beside-a-complete-structure case that union fusion exposes.

Filtering the prior rather than the candidate masks is a deliberate choice. Discarding an implausible segmentation after propagation treats the symptom; rejecting the prior that produced it removes the cause and saves the propagation. \Cref{sec:ablation} shows that this is where the available gain lies, and that aggregating candidates more cleverly is not a substitute.

\subsection{Multi-Label Merging}
The accepted anatomy masks are written into $\hat{Y}$ in the order the anatomies are declared,
\begin{equation}
    \hat{Y}(\mathbf{x}) \;\leftarrow\; a \quad \text{for each } a \in A \text{ with } \hat{M}_a(\mathbf{x}) = 1,
    \label{eq:merge}
\end{equation}
so that a later structure overwrites an earlier one wherever they overlap. This last-writer rule is deterministic but performs no arbitration, which is a second reason \cref{eq:guard} is required: without it, a drifted mask silently replaces a correct label rather than merely competing with it.

\subsection{Implementation Details}
The pipeline supports isotropic resampling with nearest-neighbour back-projection after inference, but the reported configuration leaves it disabled and operates on the native voxel grid, so predictions are scored without an interpolation round trip. Intensities are windowed to $[0, 255]$ for CT and clipped at the $[0.5, 99.5]$ percentiles for MRI. The foundation model uses the SAM2.1-Tiny backbone with the \texttt{MedSAM2\_latest.pt} checkpoint. All operations run in PyTorch \texttt{inference\_mode} with \texttt{bfloat16} precision. Ball-kernel label smoothing ($r=2$) under a 5\% volume-preservation constraint is available but disabled, consistent with the finding in \cref{sec:ablation} that added post-processing hurts once the fusion strategy no longer erodes boundaries. Unless stated otherwise we use $N=3$ prompt slices and union fusion, $\tau = 1$, both selected in \cref{sec:ablation}. Both models are loaded from public checkpoints and no weight is written at any point. Cached expert priors and pinned flags make the pipeline deterministic: re-running the $30$ ablation instances under the selected configuration reproduced the ablation result exactly, the largest per-instance DSC difference being zero.

\section{Experiments and Results}
\label{sec:experiments}

The experiments are organised around three questions. Which component of the inference-time execution policy actually determines segmentation quality (\cref{sec:ablation})? Does optimising that component improve frozen models on the target domain, and does it do so without manual prompting (\cref{sec:target})? And does a policy tuned for that domain remain compatible with established public benchmarks (\cref{sec:public})?

\subsection{Multi-cohort Evaluation Set}
\begin{figure*}[t]
    \centering
    \includegraphics[width=\textwidth]{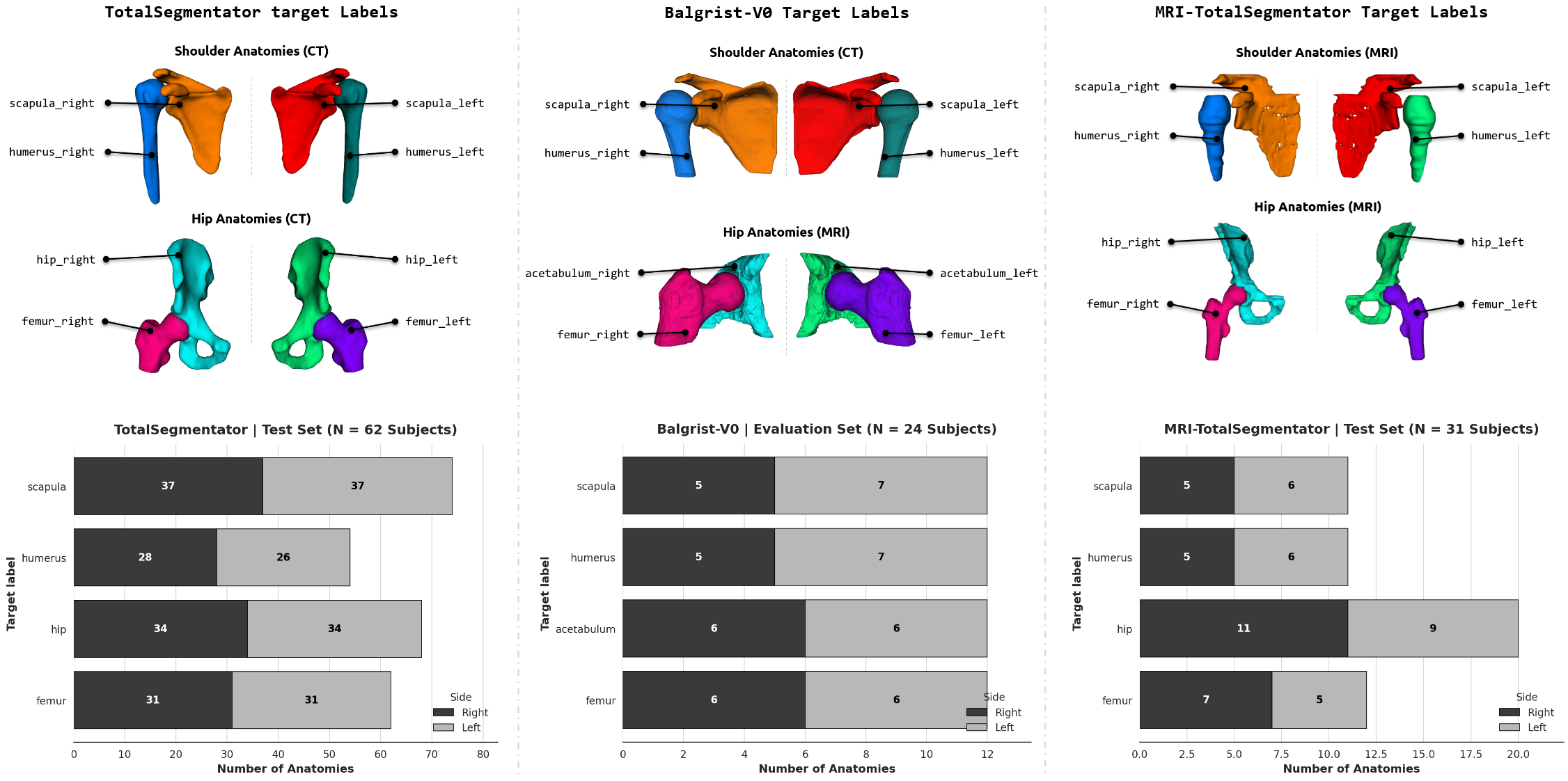}
    \caption{Multi-cohort evaluation set for hip and shoulder segmentation on CT and MRI. Each column is one cohort under a harmonised four-structure label space (scapula, humerus, hip/acetabulum, femur): the public TotalSegmentator CT test set (\emph{left}), the institutional Balgrist-V0 target domain combining PSI shoulder CT and FAI hip MRI (\emph{middle}), and the public TotalSegmentator MRI test set (\emph{right}). \emph{Top rows:} representative 3D surface renderings of the shoulder (scapula, humerus) and hip (hip/acetabulum, femur) anatomies, annotated by side (\emph{right}/\emph{left}). \emph{Bottom row:} stacked bar charts giving, per target structure, the number of segmented instances split by side, for TotalSegmentator CT ($N=62$ subjects), Balgrist-V0 CT and MRI ($N=24$ subjects; $12$ shoulder CT and $12$ hip MRI), and TotalSegmentator MRI ($N=31$ subjects).}
    \label{fig:multicohort}
\end{figure*}

We assembled a harmonised evaluation set spanning three sources with complementary imaging modalities, label protocols, and acquisition settings, covering four musculoskeletal (MSK) target structures across the hip and shoulder: the scapula, humerus, hip (os coxae)/acetabulum, and femur (\cref{fig:multicohort}). To enable cross-cohort comparison under a common label space, the TotalSegmentator ``hip'' (os coxae) bone is mapped onto the surgically relevant \emph{acetabulum} sub-region used by our institutional protocol.

The evaluation is structured as a hierarchy. The two Balgrist-V0 cohorts form the \textbf{primary evaluation}: they are the target domain, they carry independent expert manual annotations, and they are the cohorts on which the inference-time policy was selected. The two public TotalSegmentator test sets form a \textbf{secondary evaluation} of compatibility with widely used benchmarks under a policy not tuned for them.

\textbf{Balgrist-V0 CT and MRI (Evaluation Set): primary target domain.} High-resolution institutional imaging acquired for surgical planning, with independent expert manual contours as reference. It comprises $N=24$ subjects, balanced as $12$ per anatomical study and split by modality. \emph{Balgrist-V0 CT (Evaluation Set)} is the posterior shoulder instability (PSI) study, $12$ CT subjects ($5$ right, $7$ left) targeting the scapula and humerus. \emph{Balgrist-V0 MRI (Evaluation Set)} is the femoroacetabular impingement (FAI) hip study, $12$ MRI subjects ($6$ right, $6$ left) targeting the acetabulum and femoral head, with the labrum additionally contoured. Because these contours were produced independently of every evaluated model, this cohort measures cross-domain generalisation to the protocol actually used in theatre.

\textbf{TotalSegmentator CT (Test set): secondary public benchmark.} The publicly released TotalSegmentator model \cite{totalsegmentator} is an nnU-Net trained on 1204 routine CT examinations covering 104 anatomic structures. From its held-out CT test set we retained the $N=62$ subjects containing the four MSK targets, yielding scapula ($37$ right, $37$ left), humerus ($28$ right, $26$ left), hip ($34$ right, $34$ left), and femur ($31$ right, $31$ left) instances. This cohort is the in-distribution CT benchmark for the expert layer of our framework.

\textbf{TotalSegmentator MRI (Test set): secondary public benchmark.} TotalSegmentator MRI \cite{totalsegmentator_mri} extends the framework to sequence-independent MRI, training an nnU-Net on 616 MRI and 527 CT examinations covering 80 structures. From its internal MRI test set we retained the $N=31$ subjects containing the four MSK targets: scapula ($5$ right, $6$ left), humerus ($5$ right, $6$ left), hip ($11$ right, $9$ left), and femur ($7$ right, $5$ left).

\textbf{Reference standards of the public cohorts.} The reference masks of both public test sets were created through a model-assisted annotation workflow: preliminary automatic segmentations were generated and then reviewed and manually refined by expert annotators before release \cite{totalsegmentator,totalsegmentator_mri}. They are therefore expert-verified anatomical references rather than raw model predictions. They do, however, follow the same annotation protocol and the same imaging distribution on which TotalSegmentator was trained. Strong performance by that model on these two cohorts is consequently the expected outcome, and these cohorts quantify benchmark compatibility rather than behaviour under domain shift.

All inference used public checkpoints on a single workstation with an NVIDIA GeForce RTX 4090 Laptop GPU ($16$\,GB VRAM), an Intel Core Ultra 9 185H CPU ($16$ cores, $22$ logical processors), and $32$\,GB of LPDDR5 memory. A lightweight, ready-to-use implementation of the complete inference framework is publicly available at \url{https://github.com/BAL-DMU/medsam2-anatomy-inference-framework}. The raw institutional Balgrist-V0 data remains access-restricted.

\subsection{Baselines, Metrics, and Protocol}
\label{sec:protocol}
The two baselines are the frozen components of the framework, each used on its own: the anatomy-specific expert TotalSegmentator \cite{totalsegmentator} and the promptable 3D foundation model MedSAM2 \cite{medsam2}. The comparison asks whether routing a volume through both frozen models under an optimised inference policy beats querying either one directly; holding both models fixed is what attributes any difference to the policy rather than to a stronger network. Both baselines were run from their public checkpoints in their \emph{standard} released configuration: the volume is passed to the network and its output taken as-is, with no additional resampling or label-map smoothing and only the model's own internal preprocessing. MedSAM2-Anatomy uses the configuration selected in \cref{sec:ablation}. No weights are trained or fine-tuned, and no method receives a human-supplied prompt.

Performance is quantified by the Dice similarity coefficient (DSC) against the reference standard. We report the median DSC per structure because the per-instance distributions are strongly left-skewed by a small number of catastrophic baseline failures. The $95$th-percentile Hausdorff distance (HD95) and the average symmetric surface distance (ASSD) are computed alongside DSC and quoted where boundary quality is at issue. Each case carries a single expert reference, so no measured inter-reader bound is available. On the primary cohorts we therefore read DSC against the level of $0.85$ conventionally taken as clinically acceptable overlap for bone structures, and refer to it as the expert acceptance threshold; on the secondary cohorts that level has no clinical interpretation, since agreement there is with a benchmark annotation protocol rather than with contours drawn for surgical planning.

Instances were selected in three steps. The subjects of the four cohorts that carry the target labels (\cref{fig:multicohort}) provide roughly $360$ candidate anatomical instances. Instances with less than $50\%$ of the target anatomy inside the field of view were excluded manually as a quality-control step, leaving $288$. Sixteen of these were held out of the paired evaluation: fourteen because a defect in our prior-caching harness supplied seven shoulder CT scans with another subject's anatomical prior, which we report as a limitation, and two because a majority of the methods returned no mask, which indicates a problem with the scan or its reference. The evaluation therefore rests on $272$ instances, all three methods being scored on identical cases so that every comparison is paired. Where a method returns no mask for a retained instance its DSC is counted as zero; HD95 is undefined in that case and is dropped from that single median, which affects one MedSAM2 instance on the MRI hip. On TotalSegmentator CT, HD95 saturates at the $1.50$\,mm voxel-spacing floor for almost every structure and method; differences there are below the resolution of the metric, so those values are reported in \cref{tab:multicohort_dsc} without ranking.

The inference-time configuration was selected on a subset of the primary cohorts (\cref{sec:ablation}) and then applied unchanged to the complete Balgrist-V0 evaluation cohorts and to both public benchmarks. The Balgrist rows of \cref{tab:multicohort_dsc} are therefore in-sample with respect to that selection, and the two public test sets are the out-of-sample check. All results are interpreted under this hierarchy: the primary cohorts carry the evidential weight for the target-domain claim, and the secondary cohorts bound how far a policy tuned for one regime transfers to another.

\begin{table*}[!t]
\centering
\caption{Ablation study of the proposed inference-time optimization framework on a subset of the Balgrist-V0 evaluation set ($15$ subjects: $8$ CT and $7$ MRI; $30$ paired anatomical instances). Panel (a) compares alternative fusion strategies while fixing the prompt budget to $N=3$. Panel (b) evaluates the effect of the number of equidistant prompt slices while keeping the selected fusion strategy fixed. All configurations are evaluated on identical paired cases, and statistics are over per-instance DSC pooled across both cohorts. Median DSC is the primary model-selection criterion because it is less sensitive to occasional failed segmentations; $p$ is a two-sided Wilcoxon signed-rank test against the reference arm of each panel. \textbf{Bold} indicates the best value per column within a panel and \underline{underlining} the second best.}
\label{tab:ablation}
\footnotesize
\setlength{\tabcolsep}{3.5pt}
\renewcommand{\arraystretch}{0.95}
\begin{tabular}{@{}p{3.2cm}p{7.9cm}cccccc@{}}
\toprule
\multicolumn{8}{@{}l}{\emph{(a) Fusion strategy (prompt budget fixed at $N=3$)}} \\
\midrule
Fusion strategy & Description & $n$ & Mean & Median & SD & Min & $p$ \\
\midrule
Majority vote (baseline) & Majority agreement across three prompts & 30 & \underline{0.8799} & \underline{0.9238} & 0.1176 & \underline{0.5422} & ref. \\
Union & Accept voxels predicted by any prompt & 30 & 0.8683 & 0.9217 & 0.1657 & 0.2205 & 0.469 \\
Soft mean & Average prediction logits before thresholding & 30 & 0.8657 & 0.9087 & 0.1454 & 0.3393 & 0.009 \\
Weighted soft mean & Logits weighted by seed distance & 30 & 0.8344 & 0.8608 & 0.1414 & 0.4265 & 0.003 \\
Largest component & Apply connected-component filtering after fusion & 30 & \underline{0.8799} & \underline{0.9238} & 0.1176 & \underline{0.5422} & 0.587 \\
Hole filling & Add 3D hole filling after fusion & 30 & 0.8784 & 0.9209 & \underline{0.1172} & 0.5372 & 0.003 \\
\textbf{Union $+$ plausibility guard (selected)} & \textbf{Union fusion, rejecting paired anatomy below $15\%$ of its contralateral volume} & 30 & \textbf{0.9089} & \textbf{0.9371} & \textbf{0.0770} & \textbf{0.6476} & 0.058 \\
\midrule
\multicolumn{8}{@{}l}{\emph{(b) Prompt budget $N$ with the selected fusion strategy fixed}} \\
\midrule
Prompt budget & Interpretation & $n$ & Mean & Median & SD & Min & $p$ \\
\midrule
$N=1$ & Original single-prompt MedSAM2 inference; no multi-prompt fusion & -- & -- & -- & -- & -- & -- \\
$N=2$ & Not evaluated: an even prompt budget can tie under majority voting and needs an arbitrary tie-breaking rule & -- & -- & -- & -- & -- & -- \\
\textbf{$N=3$ (selected)} & \textbf{Smallest odd budget with unambiguous majority voting} & 30 & \textbf{0.9089} & \textbf{0.9371} & \underline{0.0770} & \underline{0.6476} & ref. \\
$N=4$ & Additional prompt hypothesis & 30 & 0.9001 & 0.9274 & 0.0940 & 0.5068 & 0.280 \\
$N=5$ & Larger odd prompt budget & 30 & \underline{0.9072} & \underline{0.9288} & \textbf{0.0625} & \textbf{0.7599} & 0.140 \\
\bottomrule
\multicolumn{8}{@{}p{0.99\textwidth}@{}}{\footnotesize $N=1$ denotes original single-prompt MedSAM2 and $N=2$ was omitted because majority voting can tie with an even number of prompts; dashes mark configurations not evaluated in the ablation.} \\
\end{tabular}

\end{table*}

\subsection{Which Inference-Time\texorpdfstring{\\}{ }Component Matters?}
\label{sec:ablation}
To understand which components contribute to the observed performance gain, we performed a targeted ablation on a subset of the Balgrist-V0 evaluation set comprising $15$ subjects, $8$ CT and $7$ MRI, corresponding to $30$ paired anatomical instances. This subset was the complete evaluation data available when the ablation was conducted and was used consistently across all configurations, so every comparison is paired; neither public benchmark was used for configuration selection. Because a small number of failed segmentations can substantially reduce the arithmetic mean, median DSC is the primary criterion for selecting the inference policy, while the mean, standard deviation, minimum DSC, and a two-sided Wilcoxon signed-rank test provide complementary information on robustness.

\paragraph{Fusion strategy.}
\Cref{tab:ablation}(a) compares alternative strategies for combining multiple prompt-based segmentations while fixing the prompt budget at $N=3$. Simple union fusion increases anatomical coverage, restoring the acetabulum from $0.865$ to $0.985$ of its reference volume, but it also propagates incorrect predictions, reducing the mean DSC from $0.8799$ to $0.8683$ and increasing the variability across cases. Soft-mean fusion ($p=0.009$) and weighted averaging ($p=0.003$) similarly reduce performance, indicating that voxel-wise averaging of prediction logits is less effective than combining complete anatomical hypotheses. Morphological post-processing applied after fusion yields no benefit: largest-connected-component filtering is a no-op ($p=0.587$) and hole filling is significantly harmful ($p=0.003$), because these operations address local geometric artefacts rather than incorrect anatomical priors.

The largest improvement is obtained by combining union fusion with the plausibility guard of \cref{eq:guard}, which rejects one side of a paired anatomy whenever its predicted volume is less than $15\%$ of its contralateral partner. This configuration achieves the highest mean and median DSC while substantially reducing performance variability, with the standard deviation decreasing from $0.1176$ to $0.0770$. Importantly, the improvement originates from correcting a small number of failed segmentations rather than from changing the majority of successful cases: the guard leaves $27$ of the $30$ anatomical instances unchanged and activates only when the predicted anatomy is implausibly small. On the single failure case responsible for the performance collapse, it restores the left acetabulum from $0.458$ to $0.954$ DSC and the left femur from $0.221$ to $0.944$ DSC. The guard is never activated on the public TotalSegmentator CT or MRI datasets, where both paired anatomies are consistently present, demonstrating that it does not alter normal inference. Although the improvement over the default configuration does not reach the conventional significance threshold ($p=0.058$), this configuration is selected because it systematically removes the lower tail of the performance distribution while preserving the remaining cases.

\paragraph{Prompt budget.}
\Cref{tab:ablation}(b) evaluates the effect of the number of automatically generated prompt slices while keeping the selected fusion strategy fixed. The smallest configuration, $N=1$, corresponds to the original MedSAM2 inference from a single prompt without multi-prompt aggregation and serves as the conceptual baseline. A two-prompt configuration is intentionally omitted because majority voting becomes ambiguous with an even number of hypotheses, requiring an arbitrary tie-breaking rule; only odd prompt budgets were therefore considered.

Equidistant prompt placement is motivated by our previous AssessNet-19 framework, in which sparse manual annotations on ten equidistant axial slices were sufficient to train 2D segmentation models that accurately reconstructed complete 3D anatomical volumes \cite{assessnet19}. MedSAM2-Anatomy therefore distributes its prompts uniformly along the anatomy, incorporating complementary anatomical information at low computational cost.

Increasing the prompt budget beyond $N=3$ provides little additional benefit. Although $N=5$ slightly reduces the standard deviation and increases the minimum DSC, neither larger budget significantly improves the median DSC relative to $N=3$ (Friedman $\chi^2 = 3.20$, $p = 0.20$; pairwise $p = 0.28$ and $p = 0.14$). Because computational cost increases approximately linearly with the number of prompt evaluations, from $16.6$ to $21.6$ minutes on the shoulder CT cohort, $N=3$ provides the best trade-off between accuracy, robustness, and inference efficiency and is used throughout the remainder of the paper.

These experiments demonstrate that the performance gain of MedSAM2-Anatomy is driven primarily by the inference policy rather than by increased computational effort. The most important design decision is the combination of union fusion with a lightweight plausibility guard, which selectively corrects anatomically implausible failures while leaving normal cases unchanged. Once this strategy is adopted, three equidistant prompts provide an effective balance between robustness and computational efficiency.

\begin{table*}[!t]
\centering
\caption{Quantitative segmentation performance across the four evaluation cohorts, reported per target structure as the \textbf{median Dice similarity coefficient (DSC, $\uparrow$)} and the \textbf{median $95$th-percentile Hausdorff distance (HD95 in mm, $\downarrow$)}. TotalSegmentator \cite{totalsegmentator} is the frozen expert CNN, MedSAM2 \cite{medsam2} the frozen promptable foundation model, and MedSAM2-Anatomy our training-free inference-time optimisation of the two. The upper block is the primary target-domain evaluation on the two Balgrist-V0 cohorts; the lower block is the secondary public benchmark evaluation. \textbf{Bold} indicates the best and \underline{underline} the second-best result per metric and row; tied values share a mark. All three methods are evaluated on the same $272$ paired structure instances, so every value in a row rests on identical cases. The evaluation protocol is given in \cref{sec:protocol}.}
\label{tab:multicohort_dsc}
\small
\setlength{\tabcolsep}{4.5pt}
\renewcommand{\arraystretch}{0.95}
\begin{tabular}{llcccccc}
\toprule
\multirow{2}{*}{Dataset} & \multirow{2}{*}{Structure} & \multicolumn{2}{c}{TotalSegmentator \cite{totalsegmentator}} & \multicolumn{2}{c}{MedSAM2 \cite{medsam2}} & \multicolumn{2}{c}{MedSAM2-Anatomy (Ours)} \\
\cmidrule(lr){3-4} \cmidrule(lr){5-6} \cmidrule(lr){7-8}
 & & DSC $\uparrow$ & HD95 $\downarrow$ & DSC $\uparrow$ & HD95 $\downarrow$ & DSC $\uparrow$ & HD95 $\downarrow$ \\
\midrule
\multicolumn{8}{l}{\textsc{Primary target-domain evaluation}$^{\dagger}$} \\
\midrule
\multirow{5}{*}{\shortstack[l]{\textbf{Balgrist-V0 CT} \\ \textbf{(Evaluation Set)} \\ \emph{PSI shoulder}}} & Scapula (L) & 0.8104 & \underline{15.97} & \underline{0.8777} & 16.31 & \textbf{0.8859} & \textbf{15.94} \\
 & Scapula (R) & 0.8252 & 9.78 & \textbf{0.8498} & \underline{9.24} & \underline{0.8481} & \textbf{4.50} \\
 & Humerus (L) & 0.9548 & 2.58 & \underline{0.9753} & \textbf{1.66} & \textbf{0.9791} & \underline{1.85} \\
 & Humerus (R) & 0.9498 & \underline{2.29} & \underline{0.9555} & 2.45 & \textbf{0.9735} & \textbf{2.00} \\
\cmidrule(l){2-8}
\rowcolor{gray!12} & \textbf{Cohort average} & 0.8851 & 7.65 & \underline{0.9146} & \underline{7.42} & \textbf{0.9216} & \textbf{6.07} \\
\midrule
\multirow{5}{*}{\shortstack[l]{\textbf{Balgrist-V0 MRI} \\ \textbf{(Evaluation Set)} \\ \emph{FAI hip}}} & Acetabulum (L) & 0.5096 & 34.33 & \underline{0.7334} & \underline{9.08} & \textbf{0.9119} & \textbf{4.47} \\
 & Acetabulum (R) & 0.6930 & 19.13 & \underline{0.8851} & \textbf{5.13} & \textbf{0.8909} & \underline{8.32} \\
 & Femur (L) & 0.8248 & 22.97 & \underline{0.9156} & \underline{9.56} & \textbf{0.9464} & \textbf{2.76} \\
 & Femur (R) & 0.8145 & \underline{11.74} & \underline{0.8351} & 13.28 & \textbf{0.9477} & \textbf{4.43} \\
\cmidrule(l){2-8}
\rowcolor{gray!12} & \textbf{Cohort average} & 0.7105 & 22.04 & \underline{0.8423} & \underline{9.26} & \textbf{0.9242} & \textbf{4.99} \\
\midrule
\multicolumn{8}{l}{\textsc{Secondary public benchmark evaluation}} \\
\midrule
\multirow{9}{*}{\shortstack[l]{TotalSegmentator CT \\ (Test set)$^{\ddagger}$}} & Scapula (L) & \textbf{0.9803} & 1.50 & 0.9113 & 1.50 & \underline{0.9234} & 1.50 \\
 & Scapula (R) & \textbf{0.9767} & 1.50 & 0.9215 & 1.50 & \underline{0.9312} & 1.50 \\
 & Humerus (L) & \textbf{0.9828} & 1.50 & 0.9562 & 1.50 & \underline{0.9643} & 1.50 \\
 & Humerus (R) & \textbf{0.9851} & 1.50 & 0.9590 & 1.50 & \underline{0.9652} & 1.50 \\
 & Hip (L) & \textbf{0.9908} & 1.50 & 0.9556 & 1.50 & \underline{0.9627} & 1.50 \\
 & Hip (R) & \textbf{0.9913} & 1.50 & 0.9546 & 2.12 & \underline{0.9623} & 1.50 \\
 & Femur (L) & \textbf{0.9924} & 1.50 & 0.9633 & 2.12 & \underline{0.9717} & 1.50 \\
 & Femur (R) & \textbf{0.9924} & 1.50 & 0.9661 & 1.50 & \underline{0.9724} & 1.50 \\
\cmidrule(l){2-8}
\rowcolor{gray!12} & \textbf{Cohort average} & \textbf{0.9865} & 1.50 & 0.9485 & 1.66 & \underline{0.9566} & 1.50 \\
\midrule
\multirow{9}{*}{\shortstack[l]{TotalSegmentator MRI \\ (Test set)}} & Scapula (L) & \textbf{0.6641} & 62.41 & \underline{0.5170} & \underline{45.12} & 0.4140 & \textbf{35.22} \\
 & Scapula (R) & \textbf{0.7465} & \textbf{28.44} & 0.5150 & 40.31 & \underline{0.6126} & \underline{33.56} \\
 & Humerus (L) & \textbf{0.8183} & \textbf{7.58} & 0.7095 & \underline{13.65} & \underline{0.7230} & 15.25 \\
 & Humerus (R) & \underline{0.8847} & \underline{6.47} & 0.8844 & \textbf{3.38} & \textbf{0.8922} & 9.85 \\
 & Hip (L) & \textbf{0.8888} & \textbf{3.12} & \underline{0.6445} & \underline{13.64} & 0.6216 & 20.75 \\
 & Hip (R) & \textbf{0.9003} & \textbf{3.00} & 0.7268 & \underline{11.99} & \underline{0.7903} & 12.32 \\
 & Femur (L) & \textbf{0.9632} & \textbf{1.03} & 0.9564 & \underline{2.17} & \underline{0.9585} & \underline{2.17} \\
 & Femur (R) & \textbf{0.9616} & \textbf{1.38} & \underline{0.9348} & \underline{4.50} & 0.8992 & 5.23 \\
\cmidrule(l){2-8}
\rowcolor{gray!12} & \textbf{Cohort average} & \textbf{0.8534} & \textbf{14.18} & 0.7361 & 16.84 & \underline{0.7389} & \underline{16.79} \\
\bottomrule
\multicolumn{8}{l}{\footnotesize $^{\dagger}$\,Primary target-domain cohorts with independent expert manual annotations.} \\
\multicolumn{8}{l}{\footnotesize $^{\ddagger}$\,HD95 saturates at the $1.50$\,mm voxel-spacing floor and is not ranked.} \\
\end{tabular}

\end{table*}

\subsection{Inference-Time Optimization on the Target Domain}
\label{sec:target}
The primary evaluation tests the central claim: whether an optimised inference policy over frozen models recovers segmentation quality that neither frozen model delivers alone, on imaging acquired for surgical planning and graded against independent expert annotations.

It does, and by the largest margins in the study (\cref{tab:multicohort_dsc}). MedSAM2-Anatomy is the best method on $7$ of the $8$ Balgrist structures by DSC and $6$ of $8$ by HD95, and attains the best cohort average on both metrics in both cohorts: average median DSC $0.9216$ against $0.9146$ (MedSAM2) and $0.8851$ (TotalSegmentator) on Balgrist-V0 CT, and $0.9242$ against $0.8423$ and $0.7105$ on Balgrist-V0 MRI. Boundary quality separates the methods more sharply than overlap does: average median HD95 on Balgrist-V0 MRI falls from $22.04$\,mm for the frozen expert to $9.26$\,mm for the unguided foundation model and $4.99$\,mm for the optimised composition, a factor of $4.4$ against the expert where DSC improves by a factor of only $1.3$. Since both cohorts are independent of every model's training data, this is a gain in cross-domain generalisation obtained purely by reconfiguring how frozen models are executed. Three observations support the interpretation that this is inference-time optimisation at work rather than a stronger model.

\emph{Expert priors compensate for domain shift where the expert itself cannot.} The margin widens exactly where the shift is most severe: $+0.21$ average median DSC over the frozen expert on hip MRI against $+0.04$ on shoulder CT. The expert remains a reliable localiser on the target domain even where its delineation degrades, and the framework consumes only that localisation.

\emph{Automatic prompt optimisation reduces prompt sensitivity.} The unguided foundation model, prompted without a validated prior, is unstable across sides: the right femur falls to $0.8351$ where the left reaches $0.9156$, and it returns no mask at all on one MRI hip instance, which enters as DSC $=0$. Sampling several hypotheses from a validated prior and fusing them removes that asymmetry, taking both femora above $0.94$.

\begin{figure*}[!t]
    \centering
    \includegraphics[width=0.9\textwidth]{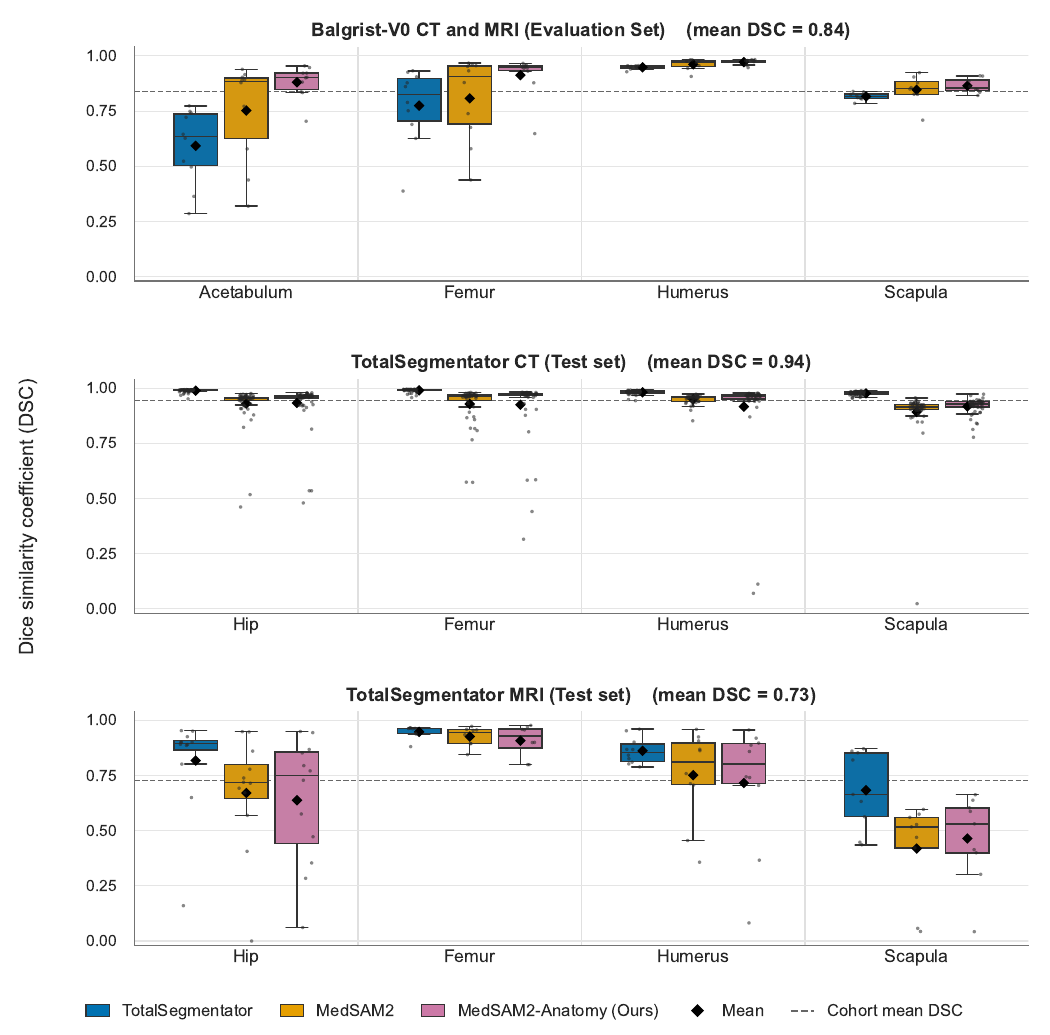}
    \caption{Per-structure DSC distributions across the primary target-domain and secondary public benchmark cohorts. Inference-time optimisation reduces low-performance outliers on Balgrist-V0, while TotalSegmentator retains the highest distributions on its public benchmarks. Rows are cohorts: the primary Balgrist-V0 evaluation sets (top) and the secondary TotalSegmentator CT and MRI test sets (middle and bottom). Boxes show the interquartile range (IQR) with the median; black diamonds indicate the mean, whiskers extend to $1.5\times$IQR, and individual cases are overlaid as points. Left and right instances are pooled. Dashed lines denote the cohort-wide mean DSC pooled over the three methods and the four structures, a within-cohort reference level rather than a quality criterion, and zero-valued points indicate cases in which a method returned no mask.}
    \label{fig:box_multicohort}
\end{figure*}

\emph{The composition is more robust than either frozen model alone.} On the target domain both baselines have a structure they fail on, and they are not the same structure. \Cref{fig:box_multicohort} shows the distributions behind the medians: for the acetabulum and femur the frozen expert is not merely lower on average but widely dispersed, its interquartile range spanning much of the scale and its lower whisker reaching far below its median, whereas the optimised composition compresses the box towards the top and largely removes that tail. Reading the Balgrist medians of \cref{tab:multicohort_dsc} against the expert acceptance threshold, both promptable methods clear DSC $=0.85$ on all four structures whereas TotalSegmentator clears it only on the humerus ($0.9508$); MedSAM2-Anatomy holds the larger margin above the threshold at every structure.

\subsubsection*{Two Regimes Within the Target Domain}
The two studies probe opposite regimes and expose the mechanism behind the aggregates: a severe domain shift with thin, low-contrast targets in the hip MRI study, against high bone contrast and a near-ceiling baseline in the shoulder CT study.

\textbf{FAI hip (MRI).} This is the most demanding setting in the benchmark and the one for which the framework was built. The frozen expert produces block-like boundaries, and its failure concentrates in the acetabulum, whose left side collapses to $0.5096$; the optimised composition restores it to $0.9119$, the largest single-structure gain anywhere in the study. Prompting a foundation model recovers much of the deficit without any learned anatomical prior, but only the validated prior makes that recovery consistent across sides. Surface error improves by more than overlap alone suggests, median ASSD falling from $3.84$\,mm for the frozen expert to $1.81$\,mm for the unguided foundation model and $1.04$\,mm for the optimised composition. The top row of \cref{fig:qualitative_multicohort} shows one such scan.

\textbf{PSI shoulder (CT).} Here bone contrast is high and the frozen expert is already near its operating ceiling, so the available headroom is small and concentrated in one structure. Almost all of the gain comes from the scapula, whose left side improves from $0.8104$ to $0.8859$, whereas the humerus, already well segmented by every method, moves only from $0.9548$ to $0.9791$; on the right scapula the optimised composition and MedSAM2 are effectively tied ($0.8481$ against $0.8498$). The pattern is structure-selective: the residual error sits in the thin scapular blade and the glenoid cavity, where the frozen expert's boundaries are least reliable, whereas the humeral shaft is already correct under every method. Boundary metrics agree and by a smaller margin, consistent with a near-ceiling baseline: median HD95 improves from $7.65$\,mm to $6.07$\,mm and median ASSD from $1.20$\,mm to $0.89$\,mm.

Read together, the two studies show that the benefit of inference-time optimisation scales with the degradation of the frozen expert and is never negative where that expert is already accurate. The one structure that remains close to the acceptance threshold is the scapula ($0.8548$), which is where the residual error of the framework concentrates.

\subsection{Behaviour Outside the Target Domain}
\label{sec:public}
The two secondary cohorts ask the complementary question: what happens when the expert prior is not the weak link. There, TotalSegmentator operates close to both the imaging distribution and the annotation protocol it was trained on, and it leads by a substantial margin, reaching average median DSC $0.9865$ on TotalSegmentator CT and $0.8534$ on TotalSegmentator MRI, ahead of MedSAM2 ($0.9485$, $0.7361$) and MedSAM2-Anatomy ($0.9566$, $0.7389$). This is the expected outcome rather than a defect of inference-time optimisation, and it is what a policy tuned for a different regime predicts: union fusion with prior rejection is designed for the case where the prior localises well but delineates poorly, so it deliberately discards the expert's boundaries. Where those boundaries are the most accurate estimate available, discarding them costs accuracy.

The informative comparison on these cohorts is therefore between the two promptable methods, both of which are out of distribution in the same way. Automatic prompt optimisation improves the frozen foundation model on both public cohorts, by $+0.0081$ average median DSC on CT and $+0.0028$ on MRI, and recovers structure-level failures such as the right scapula ($0.5150 \rightarrow 0.6126$). The improvement is small but consistent in sign, which is what the mechanism predicts when prompts are already reasonable. Read as a benchmark-compatibility check, the result is that the target-domain policy transfers without collapsing: it remains at least as good as the unguided foundation model everywhere while giving up the lead to the expert on the expert's own benchmark. \Cref{fig:box_multicohort} also tempers the MRI result: each structure there is scored on only eight to twelve instances even after pooling both sides, so the ranking rests on distributions too sparse to separate the methods reliably.

\begin{figure*}[!t]
    \centering
    \includegraphics[width=\textwidth]{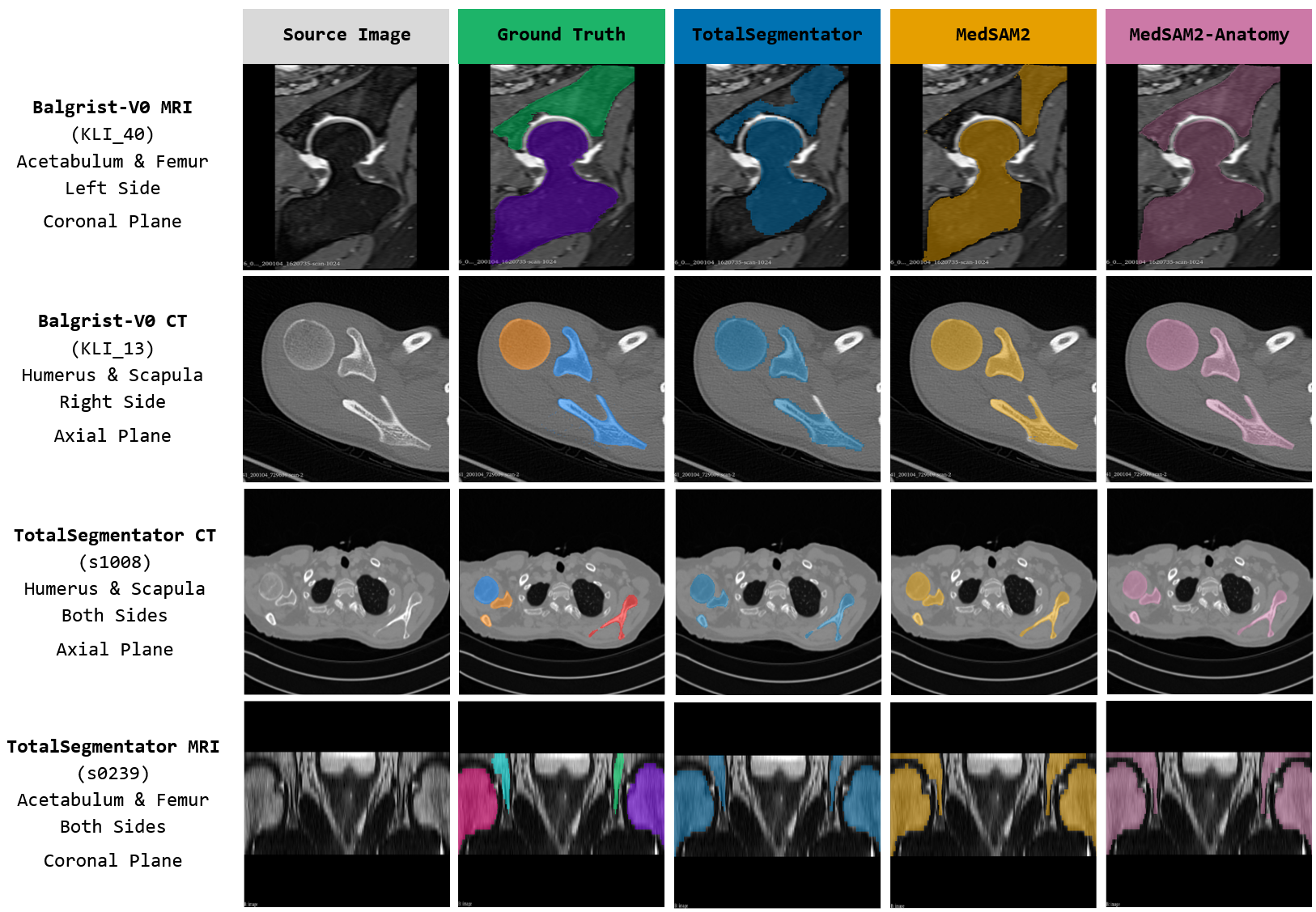}
    \caption{Qualitative comparison across the four evaluation cohorts, one subject per row. Columns show the source image, the reference annotation, and the predictions of TotalSegmentator, MedSAM2, and MedSAM2-Anatomy. The reference column colours each structure instance separately, whereas each prediction column uses that method's single colour. \emph{Row $1$} illustrates the partial-field Balgrist-V0 MRI case used to demonstrate prompt drift and its correction by the plausibility filter of \cref{eq:guard}. \emph{Rows $2$--$4$} show representative cases selected by a predefined median-based criterion, namely the instance whose three per-method DSC values deviate least, in the worst case over the three methods, from that cohort's per-method medians, for Balgrist-V0 CT, TotalSegmentator CT (ranked over the shoulder structures alone, where that cohort's residual error concentrates), and TotalSegmentator MRI. The examples illustrate typical qualitative behaviour across target-domain and public benchmark settings; the quantitative conclusions rest on \cref{tab:multicohort_dsc,fig:box_multicohort}.}
    \label{fig:qualitative_multicohort}
\end{figure*}

Taken together, \cref{sec:target,sec:public} give one result with two signs: the best inference-time policy is not fixed but depends on how reliable the expert prior is. MedSAM2-Anatomy is strongest on the two Balgrist-V0 cohorts, whose annotations were produced independently of every evaluated model, which is the target-domain generalisation that inference-time optimisation is meant to deliver; TotalSegmentator remains strongest on the two public benchmarks, which lie closest to its own training distribution and annotation protocol. The primary evidence therefore rests on the independent Balgrist-V0 evaluation, with the public benchmarks establishing compatibility rather than superiority. \Cref{fig:qualitative_multicohort} shows this case by case: its representative rows illustrate the typical outcome of each regime, and its first row separates two claims that the aggregates conflate, since on the partial-field hip MRI prompting a foundation model without a validated prior is not merely less accurate but qualitatively wrong, the track drifting across the joint to DSC $0.4375$ on the left acetabulum, whereas the plausibility filter returns the same architecture to $0.9545$.

\section{Discussion}
\label{sec:discussion}

\subsection{Why the Two Frozen Model Families Fail Differently?}
A frozen CNN encodes the intensity statistics, resolution, and field of view of its training data in its weights. Under domain shift these assumptions no longer hold at the level of fine detail, but they largely survive at the level of gross spatial layout: the network still finds the femur, it just no longer traces its surface. Our measurements match that description. On the target-domain hip MRI the frozen expert reaches average median DSC $0.7105$ with median HD95 of $22.04$\,mm and block-like boundaries, yet its localisation is good enough that prompts derived from it recover DSC $0.9242$. Domain shift degrades delineation faster than it degrades localisation, and that asymmetry is what makes an expert prior worth using even where the expert's own output is not.

Promptable foundation models have the complementary property. They delineate well because they were trained on a very broad distribution of boundaries, but they carry no anatomical prior that tells them which boundary was intended. The prompt supplies that information, and in 3D the prompt is under-specified: a seed slice fixes the appearance of a structure at one cross-section and lets memory propagation extrapolate the rest. The consequence is visible as laterality asymmetry in our results, where the unguided model reaches $0.9156$ on the left femur and $0.8351$ on the right on the same cohort with the same weights. Prompt sensitivity is not noise to be averaged away; it is the model asking a question the input does not answer.

\subsection{Why Inference-Time Optimization\texorpdfstring{\\}{ }Combines Them?}
Composing the two frozen models at inference time addresses each weakness with the other's strength, and it does so without touching either model. The expert answers the question the foundation model cannot, namely where the structure is. The foundation model answers the question the expert answers badly under shift, namely where its surface lies. Because the coupling is a routing decision rather than a learned mapping, it can be tuned on a small target-domain cohort at no annotation cost beyond what evaluation already requires, and it inherits the reproducibility of its components. With cached expert priors and pinned flags, re-running the selected configuration reproduced every per-instance DSC exactly, so reported differences between arms reflect the configuration and not run-to-run variation.

The ablation locates the tunable value precisely. Requiring agreement between independent propagations erodes exactly the structures that are only a few voxels thick: under majority voting the acetabulum is retained at $0.865$ of its reference volume against $0.985$ under union. Retaining every voxel any pass observed removes that deficit, while the outlier masks a vote was meant to suppress are better handled at their source, by rejecting an implausible prior before prompting rather than discarding a mask after it. A corollary is that post-hoc repair became counterproductive: hole filling, introduced to patch the very gaps voting had created, significantly hurt once the erosion was gone. This is a useful negative result for a literature in which such clean-up steps are commonly stacked by default.

\subsection{The Fusion Rule Depends on Prior Reliability}
The most consequential finding is not that union fusion is universally optimal, but that the optimal fusion strategy depends on the reliability of the expert prior. Union fusion with prior rejection is the right choice when the prior localises well but delineates poorly, which is the situation on high-resolution target-domain CT and MRI, where resolution and protocol differ substantially from the expert's training data. It is the wrong choice when the prior is itself the most accurate estimate available, as on the public test sets, where the expert operates near its training distribution and under the annotation protocol its training data follows. There, discarding the expert's boundaries costs accuracy, and no amount of prompt optimisation recovers it. The selected configuration should therefore be read as optimal for the Balgrist-V0 target domain, not as a general recommendation.

The current framework treats the prior as a binary object: it is either plausible enough to generate prompts or it is rejected. Nothing in the pipeline expresses the intermediate case of a prior that is partially trustworthy. The direction we consider most promising is therefore confidence-aware adaptive fusion, which decides per structure and per case whether to reject, down-weight, or fully include the expert prior. The reliability estimate driving that decision can be formed from signals available at inference time, such as agreement between the prior and the propagated candidates, prior-to-candidate volume ratios, boundary agreement, or the spread across the $N$ hypotheses, and $M^{exp}_a$ would then enter \cref{eq:fuse} as an additional term weighted by that estimate. The framework would interpolate between the two regimes we observed, remaining refinement-dominant when the prior is unreliable and becoming expert-supported when it is not, rather than committing to one behaviour in advance. We note that this is a different proposal from the soft-fusion variants tested in \cref{tab:ablation}, which weighted candidate masks by their distance from a seed slice and never estimated the reliability of the prior itself; those variants lost, and they do not bound what a confidence-aware treatment of the prior could achieve.

\subsection{Computational Trade-offs}
Inference-time optimisation buys accuracy with computation. The expert runs once per scan, but the foundation model runs $N$ times per anatomy, so cost scales as $\mathcal{O}(N)$ per structure and the framework is roughly $N$ times more expensive than querying the foundation model directly, plus one expert pass. The ablation makes the trade-off concrete and favourable: $N=5$ took $21.6$ minutes against $16.6$ for the shoulder CT cohort and returned no accuracy for the difference, so $N=3$ is selected and the overhead is bounded. Configuration selection itself is a one-off cost on a target-domain cohort, not a per-case cost. Since no gradients are computed and no optimiser state is held, memory requirements are those of inference alone, and the pipeline runs on a single GPU.

\subsection{Limitations}
Every case carries a single expert reference, so the comparison cannot be placed against measured inter-reader variability and rests instead on a conventional acceptability threshold. The benchmark covers two anatomical regions across CT and MRI, and we make no claim beyond that scope. The two public cohorts follow the imaging distribution and annotation protocol of the expert's training data, so they quantify benchmark compatibility rather than robustness under domain shift; only the two Balgrist-V0 cohorts carry annotations produced independently of every evaluated model, and those are also the cohorts on which the fusion strategy and prompt budget were selected, which makes them in-sample with respect to that choice. The selection set comprises $30$ structure instances, and the margin of the selected configuration over the default does not reach significance. A defect in our own prior-caching harness supplied seven public CT scans with another subject's anatomical prior, costing $14$ structure instances that we excluded rather than charge against any method; regenerating those priors is straightforward and remains outstanding. The labrum, although contoured in the hip reference standard, lies outside the shared four-structure label space and is not produced by any of the three methods, so we make no quantitative claim about it. Finally, the results characterise segmentation accuracy on retrospective data; they do not establish clinical readiness, which would require prospective evaluation against the downstream planning measurements the segmentations are intended to support.

\subsection{Future Directions}
The confidence-aware fusion described above is the most direct extension and the one we consider most likely to improve cross-domain robustness. Beyond it, the framework is agnostic to its components: either frozen model can be replaced without changing the surrounding logic, so a stronger expert or a stronger foundation model transfers immediately, and adding an anatomy requires only declaring an expert and a label. Prompt generation is likewise open, and we plan to explore text conditioning through models such as VoxTell \cite{voxtell} and MedGemma \cite{medgemma} to constrain the foundation model in complex multi-label settings where a mask prior alone is ambiguous. More broadly, the question this work leaves open is how far the inference-time budget can be pushed: we tuned three discrete choices on a small cohort, and the search space of routing policies over frozen models is considerably larger.

\section{Conclusion}
\label{sec:conclusion}
We asked how frozen state-of-the-art segmentation models can be systematically optimised at inference time, without retraining and without human interaction. The evidence presented here indicates that they can be, within a defined regime. MedSAM2-Anatomy optimises the execution policy of frozen models rather than their parameters: it generates prompts automatically from an expert prior, propagates several prompt hypotheses through a frozen promptable 3D foundation model, fuses the resulting candidates, and rejects anatomically implausible priors before they contribute. No weight is updated and no prompt is supplied by a user. Holding representative expert and foundation models fixed is what allows the observed differences to be attributed to that policy rather than to a stronger network. An ablation on the primary target-domain cohorts identified the fusion layer as the component that governs segmentation quality and selected union fusion with prior rejection over softer aggregation, additional prompts, and post-hoc mask repair.

On Balgrist-V0, the primary evaluation and the only cohorts whose contours were produced independently of every evaluated model, this inference-time policy raised average median DSC from $0.7105$ to $0.9242$ on FAI hip MRI and from $0.8851$ to $0.9216$ on PSI shoulder CT relative to the frozen expert, and cut median HD95 on hip MRI from $22.04$\,mm to $4.99$\,mm, while eliminating manual prompting. This is improved cross-domain generalisation obtained without touching a single weight. On the secondary public test sets, where the expert operates close to its own training distribution and annotation protocol, the same policy is not the best choice, because it deliberately discards expert boundaries that are there the most reliable estimate available; automatic prompt optimisation nevertheless still improves the frozen foundation model on both.

The conclusion is therefore not that MedSAM2-Anatomy is a better segmentation model, but that the optimal inference policy depends on how reliable the expert prior is. Making that dependence explicit through confidence-aware adaptive fusion, which decides per case whether to reject, down-weight, or include the expert prior, is the main direction we see for extending the approach to robust cross-domain automatic segmentation. These results demonstrate that training-free inference-time optimisation is a practical strategy for systematically improving frozen segmentation models under domain shift, without retraining and without manual interaction.

\bibliographystyle{IEEEtran}
\bibliography{references}
\end{document}